\documentclass{llncs} 

\usepackage[dvipsnames]{xcolor}
\usepackage{svg} 
\usepackage{subcaption}
\usepackage{comment}
\usepackage{amsmath,esvect}
\usepackage{tabularx} 
\usepackage{enumitem}
\usepackage[toc,page]{appendix}
\setlist{nolistsep}

\usepackage{graphicx} 
\newtheorem{insight}{Insight}
\newcommand{\ignore}[1]{}
\begin{document}

\title{Characterizing the Evolution of Psychological Factors Exploited by Malicious Emails}

\author{Theodore Longtchi 
 \and 
Shouhuai Xu
}

\institute{
Department of Computer Science\\ 
University of Colorado Colorado Springs\\
Colorado Springs, Colorado, USA
}


\maketitle

\begin{abstract} 
Cyber attacks, including cyber social engineering attacks, such as malicious emails, are always evolving with time. Thus, it is important to understand their evolution.
In this paper we characterize the evolution of malicious emails through the lens of Psychological Factors (PFs), which are humans' psychological attributes that can be exploited by malicious emails (i.e., attackers who send them). For this purpose, we propose a methodology and apply it to conduct a case study on 1,260 malicious emails over a span of 21 years (2004-2024). Our findings include: (i) attackers have been constantly seeking to exploit many PFs, especially the ones that reflect human traits; (ii) attackers have been increasingly exploiting 9 PFs and mostly in an implicit or stealthy fashion; (iii) some PFs are often exploited together. 
These insights shed light on how to design future defenses against malicious emails.
\end{abstract}

\keywords{Psychological Factors  \and Malicious emails \and Cyber social engineering attack \and Cybersecurity \and Impulsivity \and Trust \and Curiosity \and Defenselessness}

\section{Introduction}
Cyber social engineering attacks keep increasing despite efforts at defending against them. For example, phishing is perhaps the most prolific cyber social engineering attack \cite{trends2021apwg,zieni2023phishing,longtchi2024internet,montanez2022csekc}. The 2023 Anti-Phishing Working Group (APWG) report \cite{APWG2023report} states that 2023 has been the worst on record so far, with more than 5 million attacks, while 2022 and 2021 were respectively the record holder. 
The 2022 APWG report \cite{trends2021apwg} states that the number of phishing attacks reported to APWG has quadrupled since early 2020.
That is, the situation continues to worsen despite the endeavors at defending against phishing emails. 

A recent study \cite{longtchi2024internet} shows that attackers employ psychologically charged elements in crafting malicious emails in order to exploit human Psychological Factors (PFs), which are humans' psychological characteristics or attributes that can be exploited by cyber social engineering attackers including malicious emails (i.e., the attackers who send them). However, existing defenses rarely consider PFs \cite{longtchi2024internet,vishwanath2011people}, highlighting a mismatch between cyber social engineering attacks and defenses against them. This motivates us to investigate how PFs have been exploited by malicious emails over time, so as to shed light on designing more effective defenses to counter these attacks.

\noindent{\bf Our Contributions}.
We make three contributions. First, we reconcile the 46 PFs presented in \cite{longtchi2024internet} into 20 PFs.
This is important 
because the reconciliation would ease many tasks, such as categorizing and prioritizing PFs for defense purposes,  
and because the 46 PFs contain some overlaps.
The resulting 20 PFs can serve as a foundation or new baseline for future studies. 
Second, we propose a methodology for characterizing the evolution of PFs exploited by malicious emails over any time span and at any granularity (e.g., monthly if not weekly basis), including how to identify PFs exploited by malicious emails. 
The methodology can be adopted / adapted to identify the PFs that are exploited by other cyber social engineering attacks, or to identify other PFs (e.g., another reconciliation / refinement  of the 46 PFs \cite{longtchi2024internet} than what we propose) used by malicious emails and other cyber social engineering attacks. 
%
Third, we apply the methodology to study the evolution of the PFs that are exploited by malicious emails over the span of 21 years (2004-2024), by analyzing 1,260 malicious emails, or 60 per year, so as to draw insights into the evolution of adversary's exploitation of PFs. 
Our findings include: (i) all 20 PFs have been exploited by malicious emails during the last 21 years; (ii) attackers have been constantly seeking to exploit multiple PFs, especially the ones that reflect human traits and are dubbed Inherent PFs in this paper; (iii) attackers have been increasingly exploiting 9 PFs in a mostly implicit or stealthy fashion rather than an explicit fashion perhaps because they do not feel wise or needing to do so; (iv) some PFs are often exploited together. These insights shed light on future studies, especially the need of designing defenses to counter the exploitation of certain PFs (especially the ones that are increasingly exploited) and their collective exploitation (rather than designing defenses to counter the exploitation of individual PFs).

\noindent{\bf Ethical Issue}. We consulted with our institution's 
Internal Review Board (IRB), and were advised that we do not need the IRB's approval to conduct this study, as the dataset analyzed in this paper is provided by third parties (in addition to the 108 malicious emails from our own email box).

\noindent{\bf Related Work}. Cyber social engineering attacks have been extensively studied 
(cf., e.g., \cite{mashtalyar2021social,jampen2020don,aleroud2017phishing,longtchi2024internet,livara2022empirical,he2023method,he2024double,du2013research} and the references therein). 
However, the {\em root cause} of these attacks 
is not adequately understood despite that the notion of PF has been implicitly and occasionally explicitly mentioned as humans' weaknesses 
(e.g., \cite{parsons2019predicting}). 
The literature mainly discusses Cialdini's 6 Principles of Persuasion 
as PFs \cite{ferreira2015analysis}, often with a very small dataset (e.g., 52 emails in \cite{ferreira2015principles}). 
Few studies discuss PFs beyond these 6 Principles: {\sc impulsivity}  reflecting how fast an individual reacts to a malicious email \cite{greitzer2021experimental}, {\sc  loneliness} in the context of Covid-19 pandemic lockdown \cite{deutrom2021loneliness}, and {\sc curiosity} in phishing attacks \cite{chiew2018survey}. 
Gallo et al. \cite{gallo20212} investigate 
how phishing emails exploit cognitive vulnerabilities based on a dataset of 2-year span. Wang et al. \cite{wang2012research} study the impact of visceral triggers on phishing susceptibility, without tying to PFs.
By contrast, we consider 20 (reconciled from 46) PFs for over 21 years. In \cite{PTac-PTech-paper} we study the evolution of Psychological Techniques and Tactics exploited by malicious emails.

Not until very recently was the first systematization of PFs presented in \cite{longtchi2024internet}, including 46 PFs in 5 categories (i.e., social psychology, personality and individual differences, cognition, emotion, and workplace), which serve as a baseline for the community to scrutinize. 
Some of these PFs have been leveraged to quantify the degree of sophistication of malicious emails \cite{montanez2023quantifying}.
The present study is the first to scrutinize the 46 PFs presented in \cite{longtchi2024internet} by reconciling them into 20 PFs. We further propose a methodology to guide the identification of PFs  exploited by malicious emails, and apply it to analyze 1,260 malicious emails over a span of 21 years to characterize the evolution of PFs exploited by malicious emails. 


\noindent{\bf Paper Outline}.
Section \ref{sec:PF-refinement} presents our reconciliation of the 46 PFs proposed in \cite{longtchi2024internet} into 20 PFs. Section \ref{methodology} presents our methodology for characterizing the evolution of the PFs exploited by malicious emails.
Section \ref{case_study} applies the methodology to analyze malicious emails over any time span and at any granularity. Section \ref{limitations} discusses the limitations of the present study. Section \ref{conclusion} concludes the paper.

\section{Reconciling the 46 PFs Presented in \cite{longtchi2024internet} into 20 PFs}
\label{sec:PF-refinement}

We reconcile the 46 PFs 
to ease our analysis and provide more effective guidance on designing future defenses. This is important because it is neither wise nor practical to design defenses against the exploitation of each individual PF.

\subsection{Designing Rules to Guide the Reconciliation of PFs}

For this purpose, we propose using the following rules to guide the PF reconciliation process, with no particular precedence. 

\noindent{\bf Rule 1}: 
When reconciling a set of PFs into a single one, it is intuitive to choose the one that is {\em representative}. A representative PF is one that has been studied in a quantitative / empirical fashion, or has been studied extensively in a qualitative fashion. 
If there is still a tie, we prefer to preserving the PF that has been most exploited by attackers. Since this is a subjective matter, we use our domain knowledge to break a tie.

\noindent{\bf Rule 2}: 
In the ideal case, each PF is orthogonal or complementary to the other PFs so that there is no overlapping or redundancy among the PFs. 
Even though the 46 PFs introduced in \cite{longtchi2024internet} are carefully defined such that no two PFs have the same meaning, 
we observe that there are {\em overlaps} and {\em congruence}, as elaborated below, meaning that we need to reconcile them. (i) Some PFs may be overlapping, meaning that one PF may be replaced by multiple PFs that can collectively accommodate what is described by the particular PF. For instance, the {\sc fear} PF could be replaced by {\sc authority} and {\sc scarcity} because they each accommodates one sense of {\sc fear}.
(ii) Some PFs may be congruent, meaning that they deal with human characteristics or attributes that are inherently related to each other; as a result, one defense can, or should, cope with the exploitation of these PFs. For instance, one defense could simultaneously deal with the PFs known as {\sc authority}, {\sc respect}, and {\sc submissiveness}, suggesting that we can reconcile them into a single PF, namely {\sc authority} (per Rule 1).

\noindent{\bf Rule 3}: Assure that the input PFs (i.e., the 46 PFs in this study) are accommodated by the resulting smaller set of PFs, namely that the latter can adequately describe what the former can. 
This means that any PF that is not reconciled into, or replaced by, another PF, should be preserved.

Note that these rules are not specific to the 46 PFs we reconcile, but can be adopted / adapted to reconcile other PFs of interest.
For example, it is possible that these rules can be applied to produce another reconciliation of the 46 PFs than what is presented in this paper.


\subsection{Reconciling the 46 PFs \cite{longtchi2024internet} into 20 PFs} 

Our reconciliation  proceeds in two steps: leveraging the preceding rules to reconcile the input PFs (i.e., the 46 PFs in this study); and categorizing the resulting PFs into a small number of families. Both steps require substantial domain expertise, meaning that the result is inevitably subjective.

 
\vspace{-2em}
\begin{figure}[!htbp] 
\centering 
\includegraphics[width =.98\textwidth]{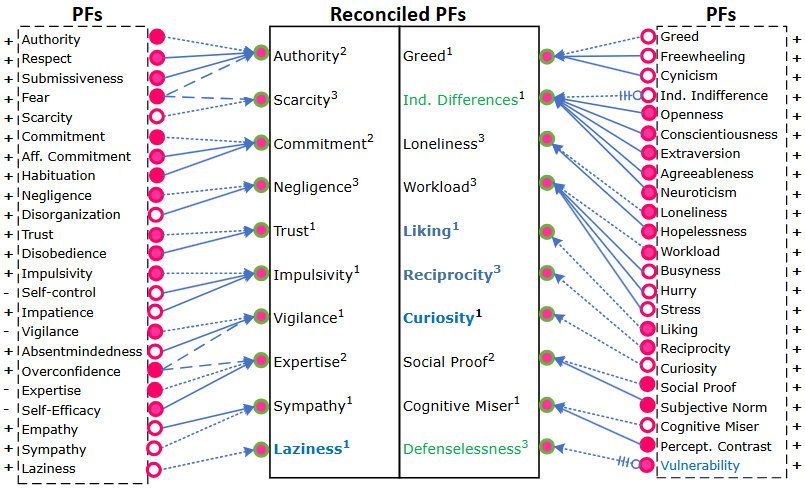}
\vspace{-0.5em}
\caption{\small Summary of the reconciliation of the 46 PFs (columns 1 and 4) into the 20 PFs, including two being renamed (in green) and 4 being preserved (in blue).
A filled circle in the original PFs (i.e., columns 1 and 4) indicates that a PF has been studied quantitatively and an empty circle indicates otherwise; the ``+'' (``-'') sign indicates that a higher PF value means a higher (lower) susceptibility to cyber social engineering attacks (e.g., malicious emails) \cite{longtchi2024internet}. ``Ind.'' is short for  Individual, and ``Percept.'' is short for perceptual. Superscripts of the reconciled PFs indicate the PF family to which it belongs (1 for Inherent PFs, 2 for Social PFs, and 3 for Situational PFs).} 
\label{fig:mapping_refined_PFs}
\end{figure}
\vspace{-2em}

\noindent{\bf Step 1: Reconciling overlapping / congruent PFs}.
Figure \ref{fig:mapping_refined_PFs} summarizes how the 46 PFs are reconciled into the 20 PFs, where a solid arrow indicates that the source PF is fully incorporated into the destination PF, a dashed arrow indicates that the source PF is partially incorporated into the destination PF (e.g., {\sc fear} is partly incorporated into {\sc authority} and partly incorporated into {\sc scarcity}), a dotted arrow indicates that the source PF is preserved while accommodating other PFs (e.g., {\sc greed}, {\sc freewheeling}, and {\sc cynicism} are reconciled into {\sc greed}) or not (i.e., {\sc liking}, {\sc reciprocity}, {\sc curiosity}, {\sc laziness}, and {\sc defenselessness} which is renamed from {\sc vulnerability} though), a dotted arrow with a tail indicates that the source PF is renamed to a destination PF (e.g., {\sc Vulnerability} renamed to {\sc defenselessness}). Details follow.

We reconcile the following four PFs {\sc authority}, {\sc respect}, {\sc submissiveness}, and {\sc fear}  \cite{longtchi2024internet} into a resulting {\sc authority} PF.
This can be justified by the observation that an effective defense against the exploitation of {\sc authority} could be effective against the exploitation of the three other PFs because {\sc authority} could demand or accommodate the three other PFs \cite{dai2022power}, while noting that {\sc authority} may imply {\sc fear} in some cultures \cite{wang2021social} and that another sense of {\sc fear} will be accommodated by the {\sc scarcity} PF below. 
The preceding Rule 1 prompts us to preserve the PF name {\sc authority} among the 4 PFs because it has been more extensively studied in a quantitative fashion than the three other PFs and because it is one of the most exploited principle of persuasion \cite{ferreira2015principles}.

We reconcile  {\sc scarcity} and {\sc fear} \cite{longtchi2024internet} into {\sc scarcity}. Although {\sc fear} of authority has been incorporated into {\sc authority} as discussed above, it is appropriate to incorporate {\sc fear} of missing out 
\cite{zhang2020fear,khetarpal2024limited} into {\sc scarcity}, which has been highly exploited by cyber social engineering attackers to make the recipient of an email to act quickly according to an attacker's proposition (e.g., ``Bitcoin price is soaring, click here to buy yours"). 
Rule 3 prompts us to preserve {\sc scarcity} because {\sc fear} has been partially incorporated into {\sc authority}.

We reconcile {\sc commitment}, {\sc consistency}, {\sc affective commitment}, and {\sc habituation}  \cite{longtchi2024internet} into {\sc commitment}, because 
these 4 PFs are congruent in the sense that they are all about one's commitment to something or one's habit of doing something 
\cite{van2019cognitive,frauenstein2020susceptibility}. Rule 1 prompts us to preserve {\sc commitment} because it is the most studied PF among these four and it is one of the most exploited principle of persuasion \cite{wang2021social,frauenstein2020susceptibility}.  

We reconcile {\sc social proof}, {\sc conformity}, and {\sc subjective norm}  \cite{longtchi2024internet} into {\sc social proof}, because 
they are congruent in the sense that they define people's behaviour according to social norms 
with respect to a group of people 
(e.g., workplace, a sport team, school). Rule 1 prompts us to preserve {\sc social Proof} because it is the most extensively studied among the three PFs (and is synonymous to {\sc conformity} per Cialdini's Principles of Persuasion \cite{frauenstein2020susceptibility}). 

We reconcile {\sc negligence} and {\sc disorganization}  \cite{longtchi2024internet} into {\sc negligence} because they are congruent in the sense that {\sc disorganization} can lead to {\sc negligence} and vice versa. Therefore, one defense that can effectively cope with the exploitation of {\sc negligence} should also be able to deal with the exploitation of {\sc disorganization}. Rule 1 prompts us to preserve {\sc negligence} because it has been studied more extensively and in a quantitative fashion \cite{ndibwile2019empirical}.  

We reconcile {\sc individual indifference} and the Big 5 personality traits which are treated as PFs---{\sc openness}, {\sc conscientiousness}, {\sc extraversion}, {\sc agreeableness}, and {\sc neuroticism}  \cite{longtchi2024internet}---into {\sc individual indifference} because  these 6 PFs are congruent in the sense that they all deal with human traits and characteristics. Thus, one defense should effectively cope with the exploitation of them. Moreover, we propose renaming {\sc individual indifference} to {\sc individual difference} 
because the latter encompasses the former and the Big 5 \cite{marengo2021meta} and because the latter
is most investigated among these 6 PFs \cite{rozgonjuk2021individual,cantillo2010thresholds}.

We reconcile {\sc trust} and {\sc disobedience} \cite{longtchi2024internet} into {\sc trust} because a breach of trust (e.g., at work) would cause {\sc disobedience}  \cite{longtchi2022sok}. Studies show individuals who are more trusting and obedient to authority are more susceptible to cyber social engineering attacks \cite{jampen2020don} and willful disobedience of employees has been exploited by cyber social engineering attacks \cite{kirlappos2014learning}. Thus, an effective defense against the exploitation of {\sc trust} would also be able to deal with the exploitation of {\sc disobedience}. Rule 1 prompts us to preserve {\sc trust} because it has been more extensively studied in a quantitative fashion than {\sc disobedience} \cite{kano2021trust}.

We reconcile {\sc impulsivity}, {\sc impatience}, and {\sc self-control}  \cite{longtchi2024internet} into {\sc impulsivity}. Note that {\sc impulsivity} and {\sc impatience} are congruent as both indicate the lack of {\sc self-control}, which makes one less susceptible to cyber social engineering attacks \cite{uebelacker2014social}. Thus, one defense that can effectively cope with the exploitation of {\sc impulsivity} could also deal with the exploitation of the other two. Rule 1 prompts us to preserve {\sc impulsivity} because 
it has been more extensively studied than {\sc impatience} and {\sc self-control}
\cite{das2019sok}. 

We reconcile {\sc vigilance}, {\sc overconfidence}, and {\sc absentmindedness}  \cite{longtchi2024internet} into {\sc vigilance} because {\sc overconfidence} may lead to lack of {\sc vigilance} and a lack of {\sc vigilance} is {\sc absentmindedness} (i.e., {\sc vigilance} is the opposite of {\sc absentmindedness}) 
while noting that a high {\sc vigilance} makes an individual less susceptible to cyber social engineering attacks. Thus, one defense that can effectively enhance {\sc vigilance} 
would address the exploitation of {\sc overconfidence} and {\sc absentmindedness} at the same time. Rule 1 prompts us to preserve {\sc vigilance} because it has been more extensively studied in a quantitative fashion than {\sc overconfidence} and {\sc absentmindedness} PF \cite{tu2019users}.

We reconcile {\sc expertise}, {\sc overconfidence}, and {\sc self-efficacy}  \cite{longtchi2024internet} into {\sc expertise} 
because it usually implies {\sc self-efficacy} (i.e., they are congruent) and because {\sc expertise} and {\sc self-efficacy} may lead to {\sc overconfidence}.
Rule 1 prompts us to preserve {\sc expertise} because it has been more extensively studied than the other two \cite{klimburg2021hacking}.

We reconcile {\sc greed}, {\sc freewheeling}, and {\sc cynicism}  \cite{longtchi2024internet} into {\sc greed} because they are congruent, namely that {\sc greed} and {\sc cynicism} involve self-interest and {\sc greed} and {\sc freewheeling} are about getting something effortlessly. This means that one effective defense against the exploitation of {\sc greed} would deal with the exploitation of the other two. Rule 1 prompts us to preserve {\sc greed} because it is more studied than the other two \cite{stajano2011understanding}.

We reconcile {\sc sympathy} and {\sc empathy} \cite{longtchi2024internet} into {\sc sympathy}; even though they have a semantic difference, both deal with feeling others' pain. Thus, they are congruent and an effective defense against the exploitation of one PF would also be effective in coping with the exploitation of the other.
Rule 1 prompts us to preserve {\sc sympathy} because it is more studied \cite{wang2021social}.

We reconcile {\sc loneliness} and {\sc hopelessness} \cite{longtchi2024internet} into {\sc loneliness} because an effective defense against the exploitation of {\sc loneliness} would also be able to deal with the exploitation of {\sc hopelessness}.
Rule 1 prompts us to preserve 
{\sc loneliness} 
because it is more extensively studied than {\sc hopelessness} \cite{deutrom2021loneliness}. 

We reconcile {\sc workload}, {\sc busyness}, {\sc hurry}, and {\sc stress} \cite{longtchi2024internet} into {\sc workload} because the other three are often congruent to, or incurred by, workload.
Rule 1 prompts us to preserve {\sc workload} 
because it is the most extensively studied among the four PFs \cite{montanez2020human}.

We reconcile {\sc cognitive miser} and {\sc perceptual contrast} into {\sc cognitive miser} because it leads to {\sc perceptual contrast}, where a person quickly compares items without thoughtfulness.  Rule 1 prompts us to preserve  {\sc cognitive miser} because it is more extensively studied \cite{mcalaney2020cybersecurity}. 

At this point, there are five PFs that have not been reconciled with others: (i) {\sc Liking}, which means that people lean favorably to, or have the desire to become, the people they like  \cite{ferreira2015principles}; (ii) {\sc reciprocity}, which means paying back for an favor that one received earlier \cite{wang2021social}; (iii) {\sc curiosity}, which is the degree of  desire to know / learn something; 
(iv) {\sc laziness}, which is the unwillingness to push pass one's comfort zone \cite{zafar2019traditional};
and (v) {\sc vulnerability}, which occurs under conditions such as failing memory or incapability of taking an informed decision. We propose renaming {\sc vulnerability} to {\sc defenselessness} to avoid potential confusions because some PFs (e.g., {\sc overconfidence}) are human vulnerabilities. The term {\sc defenselessness} 
has been used in \cite{mihelivc2021secure} 
to describe senior citizens or mentally declining people. Rule 3 prompts us to preserve these 5 PFs.  




\smallskip

\noindent{\bf Step 2: Classifying the resulting PFs into a small number of families}.
Having reconciled the 46 PFs into the 20 PFs mentioned above, now we classify the 20 PFs into a small number of families. This is important not only because the classification deepens our understanding but also because it sheds light on designing future defenses (i.e., different defenses may be required when dealing with the exploitation of different families of PFs). 

\vspace{-2em}
\begin{table}[!htbp]
\begin{tabular}{|p{2cm} | p{9.9cm}|} 
\hline
\textbf{Family} & \textbf{PFs} \\
\hline
    Inherent PFs   & {\sc liking}, {\sc individual differences}, {\sc trust}, {\sc impulsivity}, {\sc curiosity},  {\sc laziness}, {\sc vigilance}, {\sc cognitive miser}, {\sc greed}, {\sc sympathy} \\ \hline
    Social PFs   & {\sc social proof}, {\sc commitment}, {\sc authority}, {\sc expertise} \\ \hline
    Situational\newline PFs   & {\sc reciprocity}, {\sc scarcity}, {\sc negligence}, {\sc defenselessness}, \newline {\sc loneliness}, {\sc workload} \\ \hline
    \end{tabular}
    \caption{The three families of the resulting 20 PFs}
\label{tab:PF_categories}
\end{table}
\vspace{-3em}

We propose classifying the resulting 20 PFs into three families highlighted in Table \ref{tab:PF_categories}, which summarizes the superscripts in Figure \ref{fig:mapping_refined_PFs}. The three families are: Inherent PFs, which are human traits such as {\sc impatience}; 
Social PFs, which are the characteristics that humans may learn by training, such as {\sc expertise};  
Situational PFs, which are the behaviors that occur only when an individual is subjected to a situation or condition that represents an external stimulus that may cause one to act in ways to offset the stimulus, such as {\sc workload}. 



We stress that there may be other ways to classify the 20 PFs because such a classification is inevitably subjective. The one presented above is what we deem an appropriate classification.
Moreover, even for a given classification, it is still subjective to determine to which family a PF should belong. For example, we classify {\sc vigilance} into the {Inherent PFs} family even though some researchers may not treat it as a personality trait on its own but is affected by personality traits instead. This is demonstrated in an empirical study with 96 participants, which shows that {\sc vigilance} is related to (in our terminology) Situational PFs, such as {\sc workload}, but the degree of {\sc vigilance} is directly associated with the personality of a participant \cite{rose2002role}. 
As another example, we classify {\sc authority} and {\sc expertise} into Social PFs because the society creates a hierarchy among people owing to a difference in knowledge or accomplishments; as a consequence, attackers exploit the role of a high persona in the hierarchy to attack victims. 
\begin{insight}
The 20 PFs (reconciled from the 46 PFs) can serve as a new baseline for future studies on teaching and understanding cyber social engineering attacks, and possibly for guiding the design of future defenses against these attacks.
\end{insight}

\section{Methodology for Characterizing the Evolution of PFs}\label{methodology}

Given a set of PFs (such as the 20 PFs mentioned above), the objective is to use them to characterize their evolution via a set of malicious emails. For this purpose, we propose the following methodology of three steps: (i) preparing a dataset of malicious emails; (ii) identifying the PFs exploited by the malicious emails or grading the malicious emails with respect to the PFs; (iii) characterizing the evolution of PFs exploited by the malicious emails over time.

\subsection{Preparing Dataset}\label{ss:preparing_dataset}
To prepare a high-quality dataset, we propose proceeding as follows. First, one should determine the scope of a study in terms of the kinds of malicious emails, such as phishing vs. non-phishing emails and general phishing vs. spear phishing.  
Second, one should determine the time span and granularity of the malicious emails that are to be studied. It would be ideal to collect malicious emails for the longest time span possible (e.g., from the time when emails became popular) to the present time, and at the finest granularity (e.g., daily rather than monthly). However, a trade-off needs to be made in terms of the feasibility of the study, as identifying which of the 20 PFs are exploited and how by a single email is already time-consuming. 
Third, one should assure that the malicious emails under study are of high quality. This implies the following: the sources of malicious emails are trusted and credible to have a good baseline of quality data; each email must be legible, which is relevant when an email is provided in the form of image or screenshot (especially for older emails);
and each email must have all of its original content in its entirety, including logos if applicable.

\subsection{Identifying PFs Exploited by Malicious Emails}
\label{sec:grading}


For a given malicious email, we propose identifying the PFs exploited by it as follows: we grade the application of a PF as 0 for no application, 1 for implicit application, and 2 for explicit application. That is, each email will receive a score vector, where the value of each element belong to $\{0,1,2\}$. This offers us a flexibility in conducting analysis because (i) we can characterize the absence (score 0) vs. presence (scores 1 and 2) of PFs in emails and (ii) we can leverage the distinction between score 1 and score 2 to analyze implicit exploitations of PFs. Note that (ii) is important because implicit exploitations of PFs are stealthy when compared with explicit exploitations of PFs, and are perhaps harder to defend against owing to the fact that explicit exploitations could be detected by recognizing PF names in emails. Note that one email may implicitly and/or explicitly exploit multiple PFs.

To illustrate how we grade an email, which is inevitably subjective, let us look at some examples. 
If an email contains the statement ``\textit{We have updated our login system, please click here to login}'', from which we observe that PFs such as {\sc greed}, {\sc sympathy} and {\sc reciprocity} are neither implicitly nor explicitly exploited by the email, then we assign a score 0 to the email with respect to each of these three PFs. 
If an email contains the statement  ``\textit{Send it now!}'', from which we observe that {\sc impulsivity}, {\sc authority}, and {\sc workload} are (arguably) implicitly exploited (because of the word ``now''),
then we assign a score 1 to the email with respect to each of these three PFs.
If an email contains the statement ``\textit{For your commitment to staying a loyal customer, we're giving you a \$50 Walmart gift card. Click here to claim it}'', from which we observe {\sc commitment} is explicitly exploited; thus we assign a score 2 to the email with respect to {\sc commitment}. 
If an email contains the statement ``\textit{We truly appreciate your trust in our brand, see what is new}", from which we observe {\sc trust} is explicitly employed; thus, we assign a score 2 to the email with respect to {\sc trust}.
The last email is more interesting because it also implicitly exploits other PFs,
namely {\sc curiosity} and {\sc impulsivity} because of ``see what is new''; thus, we further assign a score of 1 to the email with respect to these 2 PFs, respectively.

\vspace{-1em}
\subsection{Characterizing Evolution of PFs Exploited by Malicious Emails}

Having graded a set of malicious email with respect to the 20 PFs, we can characterize the evolution of the PFs that have been exploited by malicious emails via a range of 
Research Questions (RQs), such as:
\begin{itemize}
\item RQ1: How frequently are PFs exploited by malicious emails? This allows us to understand which PFs are more exploited than others.

\item RQ2: Which PFs have been increasingly, constantly, or decreasingly exploited by malicious emails? Answering this would deepen our understanding of the trends in adversary's exploitation of PFs.

\item RQ3: Are some PFs often exploited together or all of them are exploited randomly? Answering this would shed light on designing future defenses.

\end{itemize}

\ignore{
Attackers craft emails that exploit human psychological factors as the recipients go through the emails. The objective is to lure the recipient to give in to the demand of the attacker. However, there are numerous PFs that can be exploited by attackers through well-crafted emails. This is also the case with each category of PFs (i.e., Inherent PFs, Social PFs, and Situational PFs). An attacker chooses the particular human traits or characteristics to exploit, or the social factor to exploit or the situational (or conditional) factor to exploit. Meanwhile not all attacker may be knowledgeable about targeting specific PFs, malicious emails are charged with elements that can trigger certain human traits, which is exploit by attackers. For example, an email that says "Thieves broke into my Airbnb and stole all my belongings. I need your help..." will trigger {\sc sympathy} from the recipient, whose sympathy will be exploited by the attacker if and only if the recipient follows through with the demands of the attacker. As the societal evolve Inherent PFs my also evolve, as well as social and situational factors. This implies that attackers may be evolving along side these changes, since there is evidence that social engineering attacks and breaches keep increasing. This paper characterizes the evolution of the PFs that attackers exploit via malicious emails from 2004 to 2024.  
}

\vspace{-1em}
\section{Case Study}
\label{case_study}

\subsection{Preparing Dataset}

Under the guidance of the methodology, we prepare a dataset of malicious emails as follows. First, we decide to focus on 3 kinds of malicious emails: phishing, scam, and spam. A phishing email is one that provides a file for the recipient to download or a link to click on; a scam email is one that provides a pretext to ask for the recipient's login details or that provides a narrative so that the recipient may contact the scammer; and a spam email is one that attempts to market goods or services and is considered malicious at least when the truth is hidden from the recipient or when it uses deceptive marketing (e.g., snake-oil salesman). A key difference between phishing and scam is that phishing would provide a link to click or a document to download, but scam would provide a narrative for the recipient to provide login details, possible via a different email address to get back to the sender (i.e., the scammer).

Second, we decide to collect malicious emails for the span of 2004-2024 (i.e., 21 years) and at the granularity of year.  Owing to the amount of work that is required to identify the PFs that are exploited by each malicious email, we collect 60 emails per year, leading to a dataset of 1,260 malicious emails. We collect emails from 2004 because we could not collect 60 quality emails per year for the years prior to 2004.

Third, we collect emails from the following sources: the Anti-Phishing Working Group (APWG) (250), universities or academic researchers (317), websites of tech companies and organizations (298), our own mailboxes (108), and real examples found in the Internet (287). 
The intent for considering these diversified sources (i.e., the entities that responded to our request for data) is to offer a comprehensive view of the malicious email landscape. However, this implies that the sources may have a spectrum of credibility. To address this issue, we use the same process as \cite{montanez2023quantifying}: 
we first use ScamPredictor \cite{Scamdocc46:online} to determine whether an email is phishing, spam, or scam, and then manually check each email to confirm that it is indeed malicious and correctly labelled in the three categories mentioned above (i.e., phishing, scam, and spam). We indeed observe that some emails are originally mistakenly labelled as phishing emails while they are not. 

Among the 1,260 emails, 
933 (or 74\%) are phishing emails, 305 (or 24\%) are scam emails,
and 25 (or 2\%) are spam emails. This discrepancy is inherent to the set of emails we received from the sources. 
Note that 27 (out of the 1,260) emails are originally in French (as one co-author is fluent in French). 

\subsection{Identifying PFs Exploited by the 1,260 Malicious Emails}\label{sub:identifying_PFs}

Even though we only have one grader to identify PFs from malicious emails, there may be an inconsistency in treating emails. To mitigate this inconsistency, we use Table \ref{table:PFs_Cues} 
as a reference, which is always shown on one screen of the grader. The process for identifying PFs from a single email takes about 10 minutes, meaning that grading 1,260 emails takes about 210 hours.

\begin{table}[!htbp]
{\scriptsize
\begin{tabular}{|p{2.3cm} | p{9.6cm}|} 

\hline
{PF} & {Email component from which a PF is identified} \\

\hline
\hline
{\sc Impulsivity} & ``Click {\color{blue} here} for more" / ``Special offer for you" 
\\ \hline
{\sc Trust} & ``Hey Bill, let me know if you're available" / (logos known entities)
\\ \hline
{\sc Curiosity} & ``Check my hot photos" /  ``See you credit score" 
\\ \hline
{\sc Cognitive miser} & ``We detected suspicious behaviour on your PayPal account. {\color{blue}Verify}" 
\\ \hline
{\sc Individual \newline Differences} &  ``Faculty/Personal/Student/Alumni, We received your resume application via..." / ``verify your account, and sorry for the inconvenience" 
\\ \hline
{\sc Greed} & ``You will receive 30\% of the money" /  ``Confirm your 3 million here 
\\ \hline
{\sc Liking} &   ``Protecting your account is our primary concern" 
\\ \hline
{\sc Laziness} & ``We got you covered..." / ``We offer the best price. Look no further" \\ \hline
{\sc Sympathy} & ``I'm in vacation, and I lost everything to thieves. I need your help..." 
\\ \hline
{\sc Vigilance} & ``Annazon.com" for Amazon.com. (replacing ``m" with double ``n") \\ \hline
{\sc Authority} & ``I'm Police chief..." / ``I'm Dr...." / (Use of logos such as FBI, IRS) 
\\ \hline
{\sc Commitment} & ``As a value Walmart customer, please help us take a survey..." 
\\ \hline
{\sc Social Proof} & ``I've just joined Jhoo, click here to join" (I = your idol) 
\\ \hline
{\sc Expertise} &  ``Download Covid-19 treatment procedure" (impersonating WHO) \\ \hline
{\sc {\scriptsize Defenselessness}} & ``You owe \$380 of utility bill. Make payment now" 
\\ \hline
{\sc Workload} & ``Here is the updated salary file" / ``Here is the final report"
\\ \hline
{\sc Negligence} & ``Send \$4,740 transfer immediately to XYZ for tomorrow's supply" \\ \hline
{\sc Scarcity} & ``...limited supply" / ``Stock is running out, get yours now..." 
\\ \hline
{\sc Loneliness} & `` Meet cuties in your area"  / ``Judith sent you an invite" 
\\ \hline
{\sc Reciprocity} &  ``As a former recipient yourself, please consider giving back to..."
\\ \hline
\end{tabular}
}
\caption{{\small Examples showing identification of PFs from email components.}}
\label{table:PFs_Cues}
\end{table}

\subsection{Characterizing the Evolution of the PFs}

In addressing these RQs, we deal with the presence vs. absence of a PF with respect to an email (a PF is or is not exploited). For this purpose, we treat a score 1 or 2 described in the methodology as the presence of a PF, and 0 otherwise. When the need arises, we further look at the distinction of implicit vs. explicit exploitation of a PF (i.e., score 1 vs. 2).

\noindent{\bf Addressing RQ1: How frequently exploited are PFs by malicious emails?}
In total, the 20 PFs are exploited for 4,989 times (or instances) by the 1,260 emails, meaning that one emails can exploit multiple PFs. 

\vspace{-2em}
\begin{figure}[!htbp]
\centering
\begin{subfigure}[t]{0.48\textwidth}
  \centering
\includegraphics[height=0.62\textwidth]{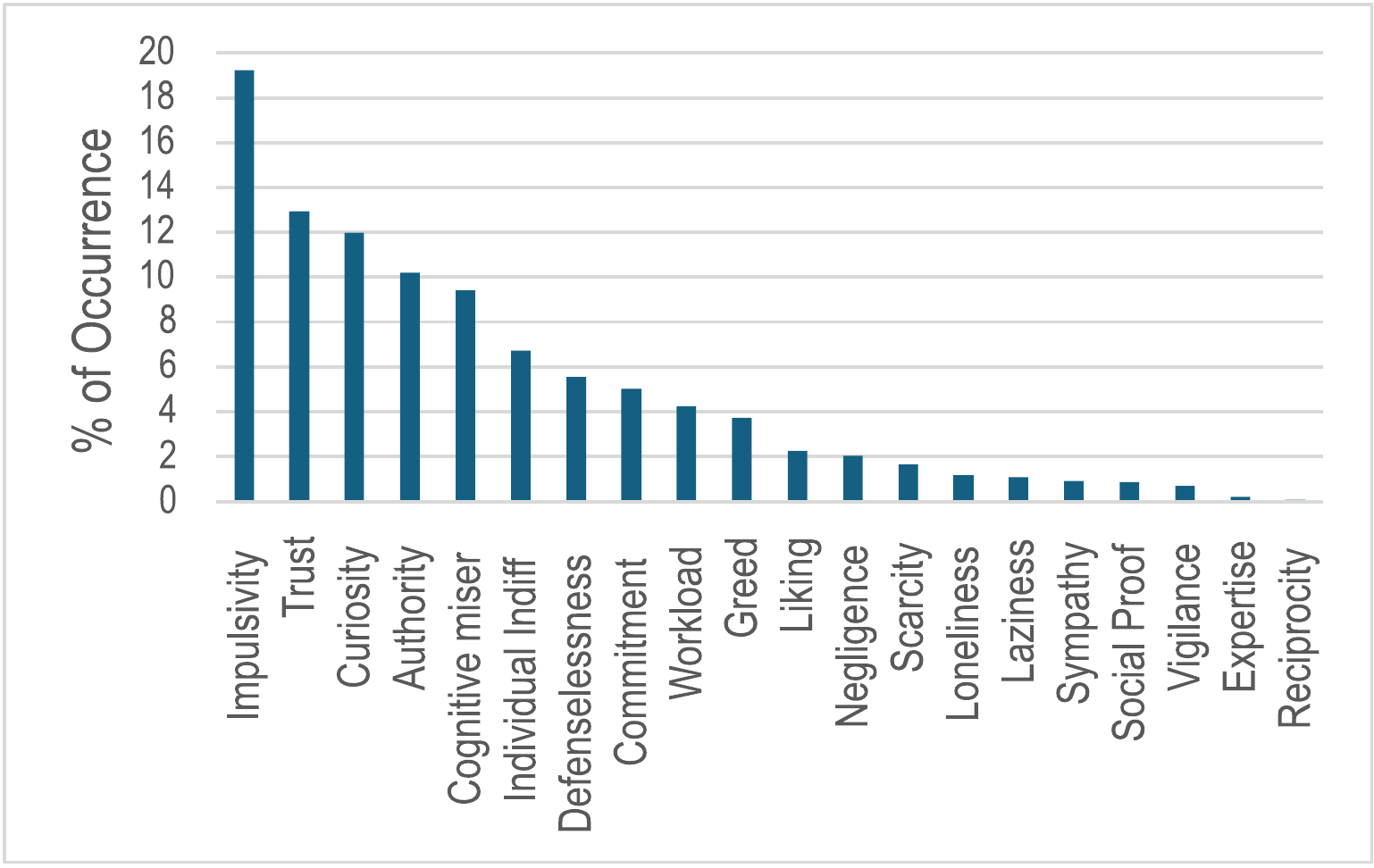}
  \caption{Occurrence of individual PFs}
  \label{fig:PFs_for_21years}
\end{subfigure}
\begin{subfigure}[t]{0.48\textwidth}
  \centering
\includegraphics[height=0.62\textwidth]{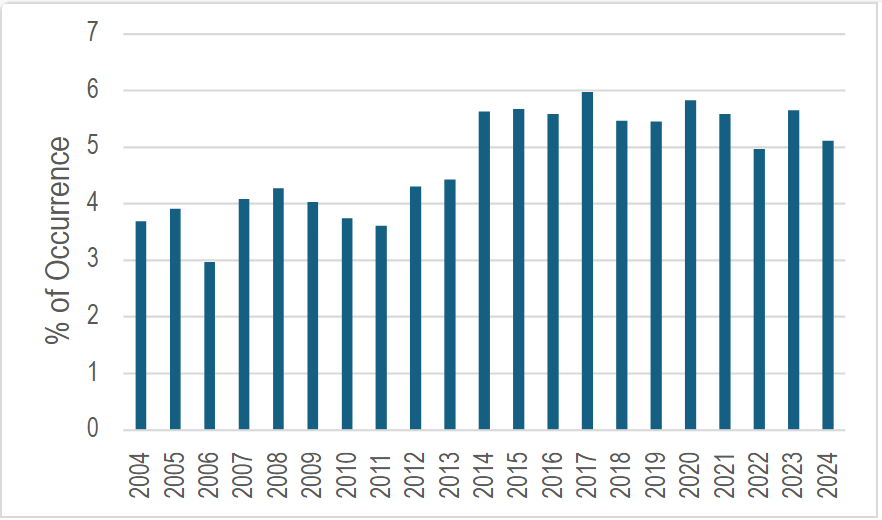}
  \caption{Occurrence of PFs per year.}
  \label{fig:PFs_by-year}
\end{subfigure}
\vspace{-1em}
\caption{{\small Frequency of PFs exploited by malicious email from 2004 to 2024.}}
\label{fig:PFs_occurrence}
\end{figure}
\vspace{-2em}

Figure \ref{fig:PFs_occurrence}(a) plots the frequency (i.e., occurrence) of individual PFs that are exploited by the 1,260 malicious emails, namely the ratio between the total number of instances of a PF that is exploited by the 1,260 emails (while noting that this number for each PF is upper bounded by 1,260 because a PF can only be exploited at most once by an email) and 4,989.
We make two observations. (i) All the 20  PFs have been exploited, highlighting that attackers are always identifying and exploiting PFs. (ii) The degree of exploitation of the 20 PFs vary: {\sc impulsivity} is the most exploited (959 times or 19.22\%), followed by {\sc trust} (646 or 12.95\%), {\sc curiosity} (597 or 11.97\%), {\sc authority} (508 or 10.18\%), and {\sc cognitive miser} (469 or 9.40\%).
Among these 5 most exploited PFs, {\sc authority} is a Social PF and the other 4 are Inherent PFs. The most exploited PF in the Situational PFs family is {\sc defenselessness} (278 or 5.57\%), ranked \#7 among the 20 PFs. This hints that attackers have exploited Inherent PFs, much more than Social PFs and Situational PFs. Thus, future research should prioritize on designing defenses to thwart adversary's exploitation of Inherent PFs (e.g., designing training schemes to educate people by focusing on these PFs).




Figure \ref{fig:PFs_occurrence}(b) plots the frequency (or occurrence) of PFs per year, namely the ratio between the total number of instances of the 20 PFs that are exploited by the 60 emails each year and 4,989. We observe that overall PFs are increasingly exploited by malicious emails. In particular, there are 1,948 instances of PFs on average between 2004 and 2013, and this number increases to 3,041 between 2014 and 2024. This means that attackers have been making effort at exploiting more PFs. By looking into the grades, we find that many PFs are exploited in all years, such as {\sc individual difference}, {\sc trust}, {\sc impulsivity}, {\sc curiosity}, {\sc cognitive miser}, and {\sc authority}, while noting that all these are Inherent PFs. However, there are a few PFs that have not been constantly exploited, such as {\sc vigilance} (an Inherent PF), {\sc sympathy} (an Inherent PF), {\sc expertise} (a Social PF), and {\sc reciprocity} (a Situational PF); in particular,  {\sc vigilance} and {\sc reciprocity} have not been exploited during the last 6 years (the caveat is that the sample is small, with 60 emails per year). 
This means that attackers have been constantly seeking to exploit most PFs, but do not appear to exploit the PFs that (i) make individuals less susceptible, such as {\sc vigilance} and {\sc expertise}, or (ii) are perhaps hard to exploit, such as 
{\sc reciprocity}.

\begin{insight}
\label{insight:PF-relevance}
Attackers have been constantly seeking to exploit most of the 20 PFs, especially the Inherent PFs for which a higher value makes individuals more susceptible (e.g., {\sc impulsivity} and {\sc curiosity}) but not the PFs for which a higher value makes individuals less susceptible (e.g., {\sc vigilance} and {\sc expertise}). 
\end{insight}

Insight \ref{insight:PF-relevance} suggests that future defenses should focus on dealing with the PFs that are often exploited by attackers, especially the PFs that are easy to exploit.


\vspace{-2em}
\begin{figure}[!htbp] 
\centering 
\includegraphics[width = .8\textwidth]{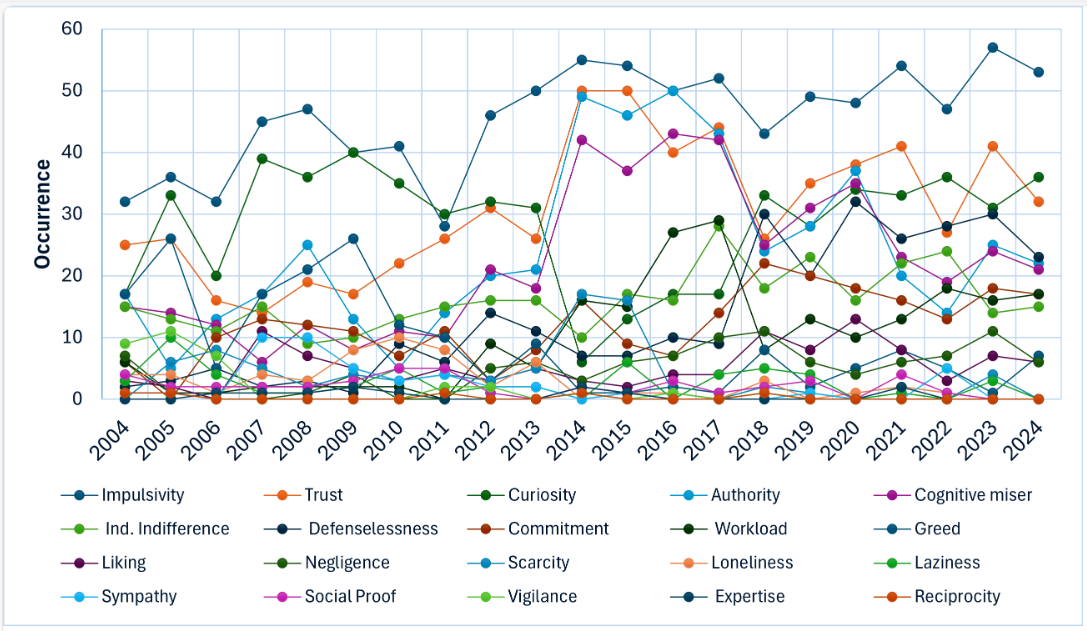}
\vspace{-1em}
\caption{\small Plots of occurrences of PFs over the 21 years where Ind. is short for Individual.}
\label{fig:individual_pfs}
\end{figure}
\vspace{-2em}

\noindent{\bf Addressing RQ2: Which PFs have been increasingly, constantly, or decreasingly exploited?}
Figure \ref{fig:individual_pfs} plots the evolution of the exploited PFs over the 21 years, where occurrence ($y$-axis) is the number of instances a PF is exploited among the 60 emails in a specific year (i.e., upped bounded by 60). Owning to the fact that there are 20 PFs, the visual effect is not perfect. However, it does show that some PFs are increasingly, roughly constantly, or decreasingly exploited. This prompts us to zoom into the detail as described below.

\vspace{-2em}
\begin{figure}[!htbp] 
\centering
\begin{subfigure}[t]{0.86\textwidth}
    \centering
\includegraphics[width = \textwidth]{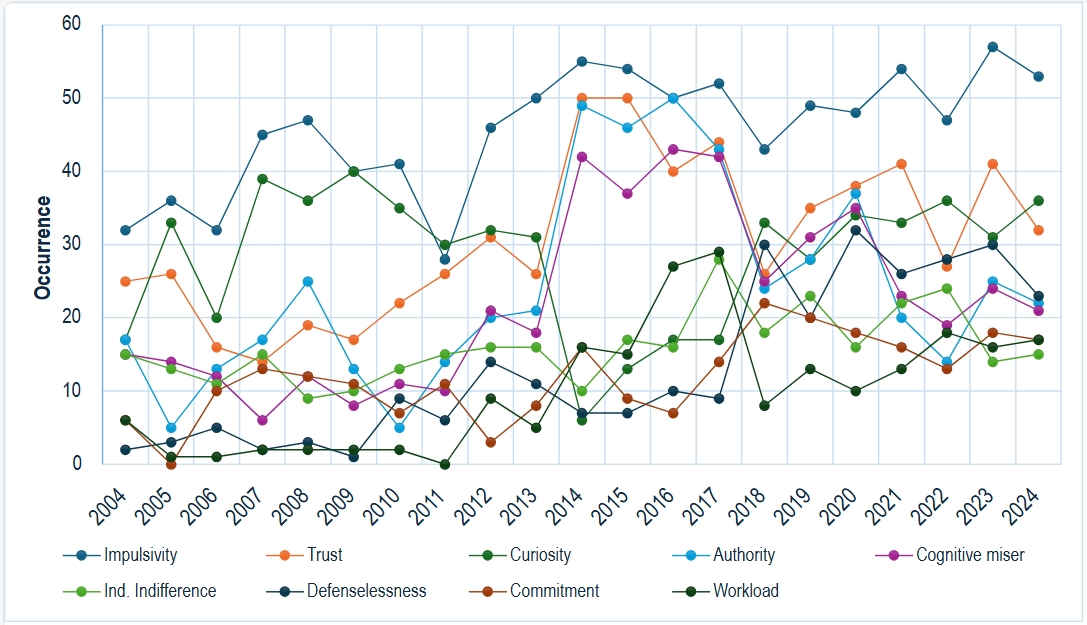}
    \caption{Plots of PFs that were increasingly exploited in the past 21 years.}
\label{fig:PFs_increasing_21years}
\end{subfigure}
\begin{subfigure}[t]{0.86\textwidth}
    \centering
\includegraphics[width = \textwidth]{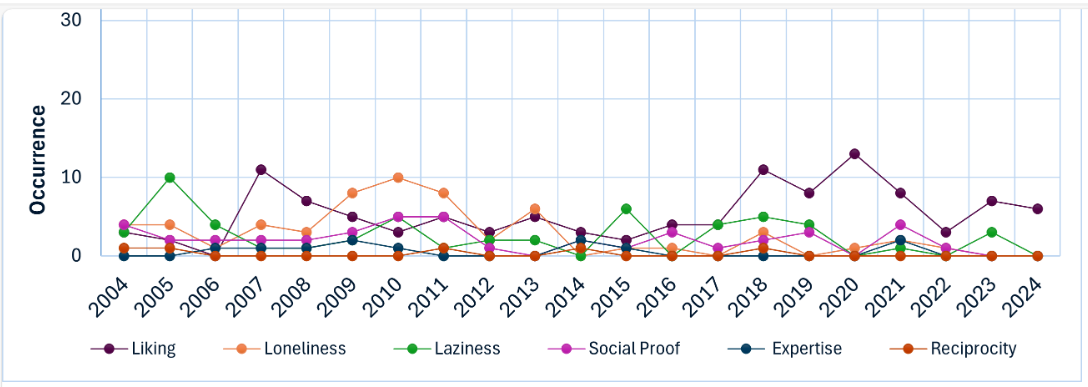}
    \caption{Plots of PFs that were constantly exploited in the past 21 years.}
\label{fig:PFs_constantly_21years}
\end{subfigure}
\begin{subfigure}[t]{0.86\textwidth}
\centering
\includegraphics[width = \textwidth]{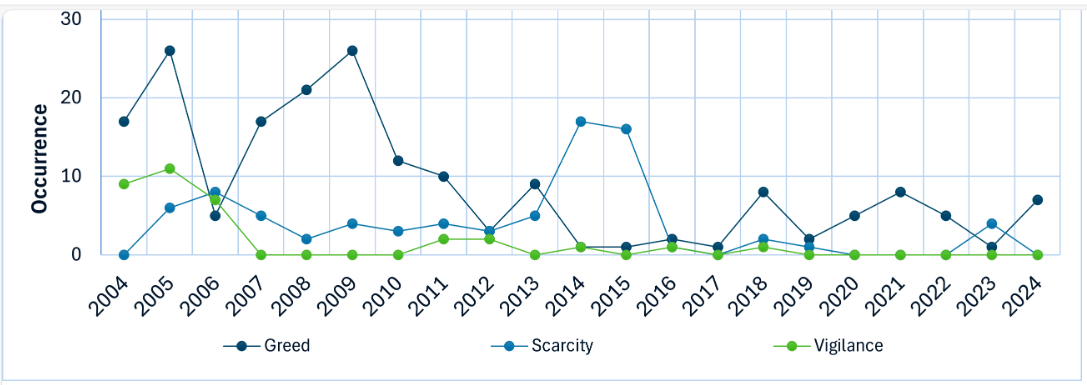}
    \caption{Plots of PFs that were decreasingly exploited in the past 21 years.}
    \label{fig:PFs_decreasing_21years}
\end{subfigure}
\vspace{-1em}
\caption{\small Plots of individual PFs, showing which PFs have been increasingly, decreasingly, and
constantly exploited by malicious emails during the 21 years.
}
\label{fig:PFs_linear_decreasing}
\end{figure}
\vspace{-2em}

Figure \ref{fig:PFs_linear_decreasing}(a)
plots the 9 PFs that have been increasingly exploited over the 21 years. 
We observe: (i) {\sc Impulsivity} (an Inherent PF) is not only the most exploited PF, but also increasingly exploited overall, suggesting that existing defenses against the exploitation of {\sc impulsivity} has not been successful. 
(ii) {\sc Defenselessness} (a Situational PF) exhibits the highest increase over the 21 years, from 2 instances in 2004 to 32 instances in 2020 before leveling to 2024.  

Figure \ref{fig:PFs_linear_decreasing}(b)
plots the 6 PFs that are, roughly speaking, constantly exploited over the 21 years. We make three observations. (i) Among these PFs, {\sc liking} and {\sc laziness} are Inherent PFs, {\sc social proof} and {\sc expertise} are Social PFs, and {\sc loneliness} and {\sc reciprocity} are Situational PFs. This means that each PF family has some PFs that are roughly constantly exploited.
(ii) These PFs are not frequently exploited (roughly speaking, they are exploited by less than 16\% of the emails), while recalling that {\sc reciprocity} (a Situational PF) is the least exploited.
(iii) When taking into account the implicit vs. explicit exploitation of a PF, we find that only 2 (out of the 9) PFs, namely {\sc trust} (an Inherent PF) and {\sc commitment} (a Social PF), are explicitly exploited by emails. This means that attackers find a way of achieving their objectives without raising suspicion. For example, an attacker would not outwardly say, ``You have to trust me, it is a legitimate deal". Instead, attackers often implicitly instill {\sc trust} in  email recipients by leveraging logos of well recognized companies / institutions or by impersonating a known personality. 

Figure \ref{fig:PFs_linear_decreasing}(c) plots the 3 PFs that have been decreasingly exploited over the 21 years, despite occasional increases. The 3 PFs include 2 Inherent PFs, {\sc greed} and {\sc vigilance}, and 1 situational PF, {\sc scarcity}.  
We make two observations. (i) The exploitation of {\sc greed} has significantly decreased from 2004 to 2024.
This may be attributed to the employment of training individuals on the consequence of being greedy when encountering malicious emails, especially scam emails such as the Nigerian Price (or 419). As a result, attackers became less interested in exploiting this PF. (ii) {\sc Vigilance} is rarely exploited perhaps because a high {\sc vigilance} indicates a low susceptibility and it is not clear how an attacker can reduce an individual's vigilance (i.e., this PF may be hard to exploit).

\begin{insight}
Attackers have been increasingly exploiting 9 PFs and mostly exploiting PFs in an implicit or stealthy fashion, suggesting that future defenses should adequately address these kinds of exploitations.
\end{insight}

\ignore{
\begin{figure}[!htbp] 
\centering 
\def\svgwidth{\columnwidth} 
\includesvg[inkscapelatex=false, width = \textwidth]{images/Greed.svg}
\caption{\small Plots of the {\sc Greed} PFs over the past 21 years, showing that {\sc greed} has the most significant decrease in exploitation in emails from 2004 to 2024.}
\label{fig:greed}
\end{figure}
}

\noindent{\bf Addressing RQ3: Are some PFs exploited together or all of them are exploited randomly?}
We use the Pearson correlation \cite{schober2018correlation}, denoted by $r$, to quantify the relationships between the 20 PFs based on the 1,260 emails (i.e., we treat the dataset as a whole without considering the years). 
\ignore{
\begin{equation*}
  r_{XY} =
  \frac{ n\sum{XY} -(\sum{X})(\sum{Y}) }{%
        \sqrt{[n\sum(X^2 - (\sum{X})^2][n\sum{Y^2}-(\sum{Y})^2}]}
    \label{eq:r}
\end{equation*}
}

\begin{figure}[!htbp] 
\centering 
\includegraphics[width = \textwidth]{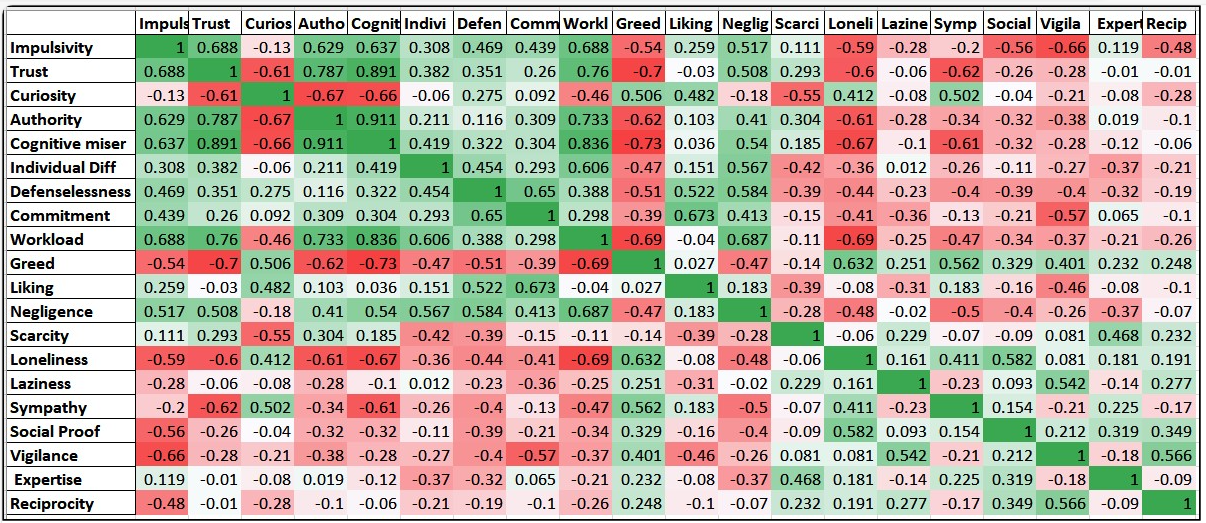}
\vspace{-1.5em}
\caption{\small Correlation between the PFs, where positive corrections are highlighted in green (with a greener color indicating a higher positive correlation) and negative corrections are highlighted in red (with a deeper red indicating a higher negative coefficient). }
\label{fig:PFs_corr}
\end{figure}

Figure \ref{fig:PFs_corr} presents the result, from which we make two observations. First, some PFs are more likely exploited together by malicious emails (i.e., positive correlation). The highest correlation is between {\sc cognitive miser} (an Inherent PF) and {\sc authority} (a Social PF), with $r = 0.91$; followed by the correlation between {\sc cognitive miser} and {\sc trust} (both are Inherent PFs), with $r = 0.89$, then by the correlation between {\sc cognitive miser} and {\sc workload} (both are Situational PFs), with $r  = 0.84$. 
Moreover, the correlation between {\sc trust} and {\sc authority} is $r=0.79$, the correlation between {\sc trust} and {\sc workload} is $r=0.76$, and the correlation between {\sc authority} and {\sc workload} is $r=0.73$. 
That is, {\sc cognitive miser}, {\sc authority}, {\sc trust}, and {\sc workload} are often exploited together by malicious emails. This sheds light on designing effective defense: A defense should cope with the PFs that are often exploited together.

\vspace{-1.5em}
\vspace{-2em}

Second, some PFs are less likely exploited together by malicious emails (i.e., negative correlation). The highest negative correlation 
is the correlation between {\sc cognitive miser} and {\sc greed}, with $r = -0.73$, followed by the correlation between {\sc trust} and {\sc greed}, with $r = -0.7$, then by the correlation between {\sc greed} and {\sc workload}, with $r = -0.69$, then by the correlation between {\sc authority} and {\sc curiosity}, with $r = -0.67$. This can be explained as follows: {\sc greed} is an action that the recipient has likely thought about the potential gains in scenarios such as the Nigerian Prince scam; in these scenarios, neither of {\sc cognitive miser}, {\sc trust}, and {\sc workload} would be relevant.
Similarly, an individual with a high {\sc curiosity} would (for example) likely click on the link presented in a malicious email without being persuaded by an entity of {\sc authority}.

\begin{insight}
While {\sc impulsivity} is the most exploited PF, {\sc cognitive miser}, {\sc authority}, {\sc trust}, and {\sc workload} are most often exploited together. This means that future defense should cope with not only the PFs it targets but also the other PFs that are often exploited together with the target PFs.
\end{insight}

\ignore{
\begin{figure}[!htbp]
\centering
\begin{subfigure}[t]{.96\textwidth}
  \centering
\includesvg[inkscapelatex=false,height=.5\textwidth]{images/Occurrence_PFs_BY2.svg}\qquad
  \caption{Plots showing the exploitation of each PF over 21 years.}
  \label{fig:individual_PFs_occurrence}
\end{subfigure}

\ignore{
\hspace{3em}
\begin{subfigure}[t]{0.45\textwidth}
  \centering
\includesvg[inkscapelatex=false, height=0.64\textwidth]{images/Occurrence_PFs_All_BY.svg}\qquad
  \caption{Unified plot showing the exploitation of all 20 PFs for 21 years.}
  \label{fig:all_PFs_occurrence}
\end{subfigure}
}
\caption{{\small The exploitation of PFs in malicious emails from 2004-2024, showing increasing exploitation from 2004 to 2024, where {\sc Impulsivity} is the most exploited PF followed by {\sc Trust} and {\sc curiosity}. (a) is modified to avoid overlaps for clarity.}}
\label{fig:Occurrence_PFs_BY}
\end{figure}
}


\section{Limitations}\label{limitations}

The present study has three limitations. First, the study is time-consuming in identifying the PFs that are exploited by malicious malicious, forcing the grading work to be done by one PhD student, meaning that the results may be biased. However, the methodology can be equally applied by multiple researchers to reduce such biases. 
Second, even though the present study is time-consuming, the sample we analyzed
(i.e., 1,260 malicious emails over 21 years span) is small compared to the large number of malicious emails on a yearly basis. Future studies need to consider a much larger sample.
Third, our refinement of the 46 PFs into the 20 PFs in three categories (i.e., Inherent, Social, and Situational PFs) might not be perfect and may oversimplify nuances between some PFs. It is an open problem to determine what would be the smallest set of PFs that would be necessary and sufficient.

\section{Conclusion}\label{conclusion}

Attackers have been exploiting human PFs (Psychological Factors) to wage cyber social engineering attacks including malicious emails.
We presented a methodology for understanding the evolution of PFs exploited by malicious emails. We conducted a case study by applying the methodology to 1,260 malicious emails during the span of 21 years (2004-2024). The case study has led to a number of useful insights, which shed light on how future defenses may be designed (e.g., defenses should adequately deal with the exploitation of the PFs that are increasingly exploited by attackers, and one defense should be able to cope with a set of PFs that are often exploited together). 

The aforementioned limitations of the present study represents exciting open problems for future research. In addition, there are at least four exciting research directions. First, it is important to study the evolution of PFs exploited by malicious websites (e.g., \cite{MirIEEECNS2022,MirIEEEISI2020Landsapce,MirIEEEISI2020MalciousWebsites,XuCNS2014,XuCodaspy13-maliciousURL}), online social networks, and other kinds of attacks. Second, it is important to study mathematical, statistical, and machine learning models to forecast the evolution of PFs, as well as PTechs and PTacs, in a fashion similar to
\cite{XuSciSec2023-forecasting,XuTIFSSparsity2021,XuGrangerCausality2020,DBLP:journals/ejisec/FangXXZ19,XuJAS2018,XuTIFSDataBreach2018,XuJAS2016,XuTechnometrics2017,XuIEEETIFS2015,XuIEEETIFS2013}. The resulting forecasting capability would allow us to design adaptive and proactive defense mechanisms (e.g., leveraging the anticipated exploitation of PFs, PTechs, and PTacs by attackers). Third, it is important to define a family of metrics to quantify the susceptibility of humans to these attacks \cite{Pendleton16,Cho16-milcom,XuSTRAM2018ACMCSUR,XuIEEETIFS2018-groundtruth,XuAgility2019,XuSciSec2021SARR}. This will not only deepen our understanding of the problem, but also offer insights into the design of effective defenses (including training schemes). Fourth, we envision that the results obtained in the preceding research directions will formulate a body of new knowledge that will become an integral component of the Cybersecurity Dynamics framework \cite{XuCybersecurityDynamicsHotSoS2014,XuBookChapterCD2019,XuMTD2020}. Cybersecurity Dynamics models can already accommodate types of cyber social engineering attacks (e.g., malicious websites-incurred drive-by download or ``pull-based'' attacks) as shown in \cite{XuTAAS2012,XuTAAS2014,XuIEEETNSE2018,XuIEEEACMToN2019,XuTNSE2021-GlobalAttractivity}. Nevertheless, systematically incorporating cyber social engineering attacks and the associated psychological aspects will lead to a more systematic body of knowledge, especially the modeling and accommodation of humans in advanced mathematical models of preventive, reactive, adaptive, proactive, and active cyber defense dynamics (see, e.g., \cite{XuTDSC2011,XuTDSC2012,XuInternetMath2012,XuGameSec13,XuHotSoS2014-MTD,XuHotSoS2015}).


\smallskip

\noindent{\bf Acknowledgement}. We thank the reviewers for their comments. This research was supported in part by NSF Grant \#2115134 and Colorado State Bill 18-086. 

\bibliographystyle{splncs04} 
\bibliography{main}

\end{document}